\documentclass[11pt,onecolumn,english,amsart,preprintnumbers,amsmath,amssymb,floatfix,nofootinbib]{revtex4}
\usepackage{tikz,xcolor}
\usepackage[colorlinks = true,
linkcolor = red,
urlcolor  = blue,
citecolor = red,
anchorcolor = blue]{hyperref}

\definecolor{lime}{HTML}{A6CE39}
\DeclareRobustCommand{\orcidicon}{%
	\begin{tikzpicture}
	\draw[lime, fill=lime] (0,0)
	circle [radius=0.16]
	node[white] {{\fontfamily{qag}\selectfont \tiny ID}};
	\draw[white, fill=white] (-0.0625,0.095)
	circle [radius=0.007];
	\end{tikzpicture}
	\hspace{-2mm}
}

\foreach \x in {A, ..., Z}{%
	\expandafter\xdef\csname orcid\x\endcsname{\noexpand\href{https://orcid.org/\csname orcidauthor\x\endcsname}{\noexpand\orcidicon}}
}

\usepackage[toc,page]{appendix}
\usepackage{epsfig}
\usepackage{graphics}
\usepackage{latexsym}
\usepackage{amsmath}
\usepackage{amssymb}
\usepackage{rotating}
\usepackage{subfigure}
\usepackage{bm}
\usepackage{color}
\usepackage{ulem}
\usepackage{booktabs}
\usepackage{caption}
\usepackage{float}
\usepackage{placeins}         
\usepackage{multirow}

\usepackage{placeins}
\usepackage{needspace}
\usepackage{xcolor}
\definecolor{color}{RGB}{0,0,0} 
\usepackage{siunitx}

\usepackage{setspace}
\newcommand{\RNum}[1]{\uppercase\expandafter{\romannumeral #1\relax}}

\begin{document}


\title {Impact of momentum-dependent drag coefficient on energy loss of charm and bottom quarks in QGP}

\author{Marjan Rahimi Nezhad$^{1}$\orcidA{}}
\email{marjanrahimi29@gmail.com}

\author{Fatemeh Taghavi-Shahri$^{1}$\orcidA{}}
\email{taghavishahri@um.ac.ir}

\author{Kurosh Javidan$^{1}$\orcidC{}}
\email{javidan@um.ac.ir}

\affiliation {
$^{(1)}$Department of Physics, Ferdowsi University of Mashhad, P.O.Box 1436, Mashhad, Iran  }
\date{\today}



\begin{abstract}\label{abstract}

This paper investigates the influence of heavy-quark momentum on their interaction rate and the resulting drag coefficient in a quark-gluon plasma.
\textcolor{color}{To go beyond simplified treatments, we introduce a phenomenological extension of the drag coefficient by expressing the energy loss coefficients as polynomial expansions of momentum, thereby providing a flexible framework to test the sensitivity of heavy-quark observables to additional momentum dependence in transport coefficients.}
Furthermore, the effects of particle momentum on radiative and collisional energy loss are determined more accurately.
The study focuses on calculating the nuclear modification factor ($R_{AA}$) of charm and bottom quarks in Pb-Pb collisions
at $\sqrt{S_{NN}} = 5.02 \: TeV$. The initial distribution functions are evolved numerically using the Fokker–Planck equation. 
The results are compared with the latest experimental data from ALICE and ATLAS, collected in 2021 and 2022.

\end{abstract}

\pacs{12.38.Bx, 12.39.-x, 14.65.Bt}

\maketitle


\section{Introduction}\label{sec:Introduction}

The main goal of ultra-relativistic heavy ion collisions at the Relativistic Heavy-Ion Collider (RHIC) and the Large Hadron Collider (LHC) is to investigate QCD matter under extreme conditions. These experiments involve the collision of heavy ions, such as gold or lead, at very high energies to create a hot, dense environment of deconfined quarks and gluons, known as the Quark-Gluon Plasma (QGP) \cite{STAR:2005gfr, PHENIX:2004vcz, QGP.Collins:1974ky}. This highly excited state of matter, whose main constituents are light quarks and gluons, exhibits properties similar to a nearly perfect fluid. Therefore, it can be effectively described using hydrodynamic models while the system reaches local thermal equilibrium \cite{Ollitrault:2007du, Becattini:2014rea}.
\\
Heavy quarks (such as charm and bottom) play an essential role in studying the properties of a quark-gluon plasma, as they are the best probe for determining the medium's transport properties \cite{HQ.vanHees:2005wb, HQ.Rapp:2009my}.
These quarks are too heavy to be significantly produced through the interaction of thermal particles in the QGP. Therefore, the dominant mechanism for their production arises from hard interactions between partons within the colliding nuclei.
Moreover, due to their large mass, these particles require a longer duration to reach thermal equilibrium with the environment. In some cases, they may even leave the plasma before achieving equilibrium.
Thus, they are valuable probes to study the entire spacetime history of the deconfined medium. 
To gain a deeper insight into the properties of heavy quarks re-scattering in the QGP, one can analyze the evolution of their distribution function over time in the transverse momentum ($P_T$) plane.
\\
Heavy quarks experience energy loss as they propagate through the hot, dense medium created by collisions.
This energy dissipation occurs through two primary mechanisms: elastic scatterings with the constituents of the QGP and gluon bremsstrahlung or radiation resulting from interaction with thermal quarks and gluons.
From a transport-theoretical perspective, the interaction between heavy quarks and the dynamically evolving QGP can be characterized by transport coefficients, which encode the cumulative effect of multiple microscopic interactions with the medium.
Among these coefficients, the drag coefficient plays a central role, as it quantifies the average resistance experienced by a parton while traversing the plasma and is directly related to its energy loss per unit path length.
\\
Extensive theoretical efforts have been devoted to obtaining more accurate estimates of the drag and diffusion coefficients. 
In particular, several studies have investigated the momentum dependence of the drag coefficient \cite{Mazumder:2011nj, Prakash:2023hfj}, as well as the momentum or temperature dependence of the diffusion coefficient \cite{Hong:2023cwl, Li:2019lex}. 
\textcolor{color}{for further examples, see also \cite{He:2011qa, Xu:2017obm, Das:2015ana, Das:2024vac}.}
In parallel, significant progress has been made in developing theoretical models for heavy-quark energy dissipation in the medium,
including both collisional \cite{Peshier:2006hi, Braaten:1991jj, Peigne:2008nd, J. D. Bjorken, Braaten:1991we} and radiative \cite{Armesto:2003jh, Armesto:2004vz, Gyulassy:2000fs,Djordjevic:2003zk} energy loss mechanisms. For a comprehensive review of many of these formalisms, see Ref. \cite{Jamil:2010te}.
\\
An accurate determination of the drag coefficient requires accounting for both the frequency of interactions experienced by heavy quarks during their propagation and the average amount of energy lost in each interaction.
As heavy quarks propagate through the thermal bath, their energy and momentum decrease, leading to a reduction in the elastic and inelastic interaction rates over time.
In this paper, we introduce a momentum-dependent extension of the drag coefficient by using a first-order polynomial ansatz applied to collisional and radiative energy-loss coefficients. This approach dynamically links the drag force to the underlying energy-loss mechanisms, 
and goes beyond simplified approaches where the drag coefficient is treated as momentum-independent \textcolor{color}{or where the momentum dependence is only that inherited from the leading-order pQCD energy-loss formulas.}

Our framework enables a more detailed assessment of how momentum-dependent scattering rates influence collisional and radiative energy loss, and how they modify the nuclear modification factor of charm and bottom quarks.
\\
It is important to clarify the phenomenological nature of this study. Our primary aim is to establish 
a baseline study that isolates the specific role of momentum dependence in the drag coefficient.
To isolate this specific effect in a controlled setting, we employ several simplifying approximations. 
The drag coefficient is parametrized using a linear momentum expansion; the medium evolution is described by a Bjorken flow with a uniform transverse profile; phenomenological Fokker–Planck transport is applied; and an effective procedure is used to compare the results with D-meson data.
These choices allow us to identify the genuine impact of the momentum-dependent terms, independently of model-specific hydrodynamic or hadronization uncertainties, providing a clear foundation for more complex and advanced studies of heavy-quark energy loss in future work.
\\
This paper is organized as follows: In the following section, we present a brief overview of the Fokker–Planck equation. We also discuss the space–time evolution of the system and specify the initial conditions employed in our analysis. 
In Section \ref{sec: DragDiffusion}, the drag and diffusion coefficients for collisional and radiative processes have been discussed. Our results are presented in Section \ref{sec:results}, and section \ref{sec: summary} is dedicated to summary and discussions.

\section{System evolution}\label{sec: Theoretical Framework}

Heavy quarks reach thermodynamic equilibrium more slowly compared to lighter particles.
Consequently, the interaction between heavy quarks and medium can be considered as interaction between equilibrium and non-equilibrium states.
The equilibrium component, known as Quark-Gluon Plasma, is presumed to form at an initial temperature ($T_i$) and time ($\tau_i$) after the nuclear collision. The QGP expands and cools due to its high internal pressure, until transition to the hadronic phase at temperature $T_c$. Heavy quarks as the non-equilibrium component, 
exhibit Brownian motion within the thermal environment of the QGP \cite{Moore:2004tg, Alam:1994sc}. 
Therefore, Fokker-Planck equation, which is a simplified version of the Boltzmann equation for small momentum transformations \cite{Svetitsky:1987gq, Rapp:2009my}, provides a suitable framework for investigating the evolution of the heavy quark's momentum distribution. According to this equation we have:
\begin{equation}
\frac{\partial}{\partial t}f(p,t) = -\frac{\partial}{\partial p_{i}}\left[A_{i}(p)f(p,t)\right] + \frac{\partial^2}{\partial p_{i}\partial p_{j}}\left[D_{ij}(p)f(p,t)\right]
\end{equation}
Where $f(p,t)$ represents the distribution function of heavy quarks 
and $A_{i}(p)$ and $D_{ij}(p)$ are related to the drag and diffusion coefficients, respectively. These coefficients characterize the interaction of heavy quarks with the QGP and are generally influenced by both temperature and momentum. 
The i and j indices represent different spatial directions, which can be omitted as we are considering a spatially uniform quark-gluon plasma in transverse plane. In the next section, we will discuss these coefficients in detail.
\\
The initial transverse momentum distribution of charm and bottom quarks has been taken from Refs. \cite{Sheibani:2021ovo, Modarres:2021gva, Olanj:2020lkt}.
In \cite{Sheibani:2021ovo}, the distribution functions are obtained using the APFEL legacy framework, incorporating the nCTEQ15 model and by considering the EMC effect to determine bound quarks within the nuclei at a specific center-of-mass energy.
Subsequently, the Pythia8 simulation code is employed to calculate the initial spectra of heavy quarks after the Pb-Pb collision. The process of converting distribution functions from x-variable to transverse-momentum-dependent PDFs can be found in relation (1) of this article \cite{Sheibani:2021ovo}.
\\
As we mentioned, the QGP environment is constantly changing and evolving, and its temperature decreases during expansion. The temperature evolution of the background (QGP) is governed by relativistic hydrodynamics. Here, we consider the evolution equation of the QGP temperature, using the Bjorken flow in the Milne coordinate, as follows \cite{Chattopadhyay:2018apf, Grozdanov:2015kqa}:
\begin{equation}\label{Eq:eq1}
T(\tau) = T_0\left(\frac{\tau_0}{\tau}\right)^{\frac{1}{3}}\left[1+\frac{2}{3\tau_0T_0}\frac{\eta}{s}\left(1-\left(\frac{\tau_0}{\tau}\right)^{\frac{2}{3}}\right)\right]
\end{equation}
Where $T_0$ and $\tau_0$ represent the initial temperature and proper time, respectively. In this article, the initial proper time has been taken as \(\tau_0 = 0.33~\text{fm}/c\) and the initial temperature has been set to $T_0=$ \SI{403}{\mega\electronvolt}. The parameter $\eta /s$ denotes the ratio of viscosity to entropy, which can be treated as a constant with a value of $1/4\pi$ \cite{Ruggieri:2013ova}.
\\
Furthermore, utilizing a temperature-dependent function for the running coupling $\alpha_s(T)$ is crucial, as temperature significantly influences the QCD coupling in the QGP system. We define $\alpha_s(T)$ as \cite{Braaten:1989kk}:
\begin{equation}\label{alphaaa}
	\alpha_s(T) = \frac{6\pi}{(33-2N_f)\ln\left(\frac{19 \: T}{\Lambda_{MS}}\right)}
\end{equation}
We take $N_f=3$ as the number of active flavors present in the QGP, with the QCD cut-off parameter set at $\Lambda_{MS}=$ \SI{80}{\mega\electronvolt}.
\\
The dominant influence of the medium's temperature on plasma evolution is encoded in the temperature dependence of the strong coupling constant, as expressed in Eq.~(\ref{alphaaa}).
However, other effects on the system dynamics arise through the temperature and momentum dependence of the drag and diffusion coefficients, which are discussed in the following section.

\section{Drag and diffusion coefficients}\label{sec: DragDiffusion} 

To solve the Fokker-Planck equation and derive the distribution function of heavy quarks at the transition temperature ($T_c$), we require the drag and diffusion coefficients as a function of system parameters. 
It is important to note that the drag coefficient carries information about the dynamics of heavy quark interactions with the medium and is influenced by the properties of the thermal bath. Consequently, the key aspect of determining the temporal evolution of heavy quark distribution function lies in calculating the drag force acting on them, or equivalently, the rate of energy loss per unit distance traveled by the heavy quark in the medium. Thus, we can write:
\begin{equation}\label{drag0}
A(p) \propto -\frac{1}{p}\frac{dE}{dL}
\end{equation}
We consider both kinds of collisional and radiative energy loss in calculating the drag coefficient. Therefore, it can be determined through the following relation:
\begin{equation}\label{1st Drag}
	A(p) = -\frac{1}{p} \:  \left[ k_1 \: (\frac{dE}{dL})_{coll}+ k_2 \: (\frac{dE}{dL})_{rad} \right]
\end{equation}
The parameters $k_1$ and $k_2$ are typically treated as constants, which can be derived from the best fit to the experimental data. These constants incorporate the contributions from collision rates, degeneracy factors, and other relevant physical factors.
\\
However, to precisely determine the drag coefficient, it is necessary to consider the impact of particles' momenta on the rate of their collisions, which directly influences energy loss.
%
%
\textcolor{color}{As a heavy quark propagates through the QGP, its momentum evolves through interactions with the medium, leading to modifications in the scattering kinematics, cross sections, and effective interaction strength of both elastic and inelastic processes. Therefore, the resulting transport coefficients inherently depend on the particle momentum.
Standard pQCD formulations already contain an intrinsic momentum dependence through the underlying scattering processes. However, in the intermediate momentum region ($p_T \sim 2$--$15$ GeV), these approaches may not fully capture all relevant physical effects.
In particular, several contributions --- including running coupling effects, higher-order corrections, non-perturbative interactions, finite-mass effects, and incomplete Landau-Pomeranchuk-Migdal (LPM) suppression --- can modify both the magnitude and the momentum dependence of the energy loss \cite{Hong:2025dfj, Xing:2021xwc, Liu:2020dlt, Ke:2018jem, Mehtar-Tani:2019ikq}. These effects are difficult to incorporate consistently within a single microscopic framework and may lead to an effective modification of the transport coefficients.
\\
Motivated by this, we introduce a phenomenological extension in which the coefficients governing collisional and radiative energy loss are allowed to acquire an additional momentum dependence through a linear expansion. This parametrization does not replace the intrinsic momentum dependence of the underlying energy-loss mechanisms, but rather provides a flexible framework to capture missing or unresolved physics in a controlled, data-driven manner. In this sense, the present approach should be viewed as a phenomenological extension analogous to data-driven extractions of transport coefficients.}
\\
Consequently, we have formulated the drag coefficient as follows:
\begin{equation}\label{draggg}
	A(p) =-\frac{1}{p} \left[ \: k' (p) \left( \frac{dE}{dL} \right)_{coll} + \: k'' (p) \left( \frac{dE}{dL} \right)_{rad} \: \right] 
\end{equation}
The functions $k'(p)$ and $k''(p)$ represent collisional and radiative energy loss coefficients, respectively. 
We assume these functions are comprised of a constant value and a modified term which is expressed as polynomial expansions of momentum, as follows:
\begin{equation}\label{bast1}
\begin{aligned}
	k' (p) \:\approx \: k_1 \: (\: 1 + \alpha \; p + O(P^2) \:)  \\
    k'' (p)\: \approx \: k_2 \: ( \: 1 + \beta \: p + O(P^2) \:)
\end{aligned}
\end{equation}
We consider these expansions up to the first-order which include linear terms.
Indeed we can ignore higher-order terms due to the lack of sufficient experimental data. 
\textcolor{color}{By treating the collisional and radiative energy-loss coefficients as independent functions of momentum, our parametrization allows for separate momentum-dependent adjustments of each energy-loss channel.}
Setting $\alpha$ and $\beta$ to zero yields the original expression of the drag coefficient in the equation (\ref{1st Drag}).\\
Finally, by substituting (\ref{bast1}) into (\ref{draggg}), we have:
\begin{equation}\label{bigdrag}
	A(p) =-\frac{1}{p} \left[ \: k_1 (1 + \alpha \; p ) \left( \frac{dE}{dL} \right)_{coll} + \: k_2 ( 1 + \beta \: p) \left( \frac{dE}{dL} \right)_{rad}\: \right] 
\end{equation}
All coefficients including $k_1$, $k_2$, $\alpha$, and $\beta$ should be determined by fitting to experimental data.
\\

Once the drag coefficient has been determined, the diffusion coefficient can be obtained via Einstein’s relation,
which is commonly applied in the weak-coupling or phenomenological context \cite{Diffusion.Srivastava:2016igg,Diffusion.Das:2010tj,Diffusion.Akamatsu:2008ge}. 
According to this relation, we have: 
\begin{equation}\label{diffu}
D(p) = T A(p) E
\end{equation}
Here, T is the temperature of the thermal bath and E is the heavy-quark energy. Therefore, by calculating the drag coefficient, we will have both the drag and diffusion coefficients.
\\
It is important to emphasize that here the Einstein relation is employed purely as a phenomenological closure within the Fokker–Planck framework, rather than as a weak-coupling assumption derived microscopically from perturbative QCD. This approach is widely adopted in strongly-coupled heavy-quark transport studies \cite{Moore:2004tg, vanHees:2004gq, Das:2010tj} and provides a consistent coarse-grained description of Brownian motion of heavy quarks in the QGP. The strong-coupling dynamics is effectively encoded in the momentum-dependent drag coefficient $A(p)$, whose functional form is constrained directly through comparison with experimental $R_{AA}$ data.
\\

In order to calculate collisional and radiative energy loss, various methods have been proposed in recent studies. In our previous work \cite{Nezhad:2023mvb}, we evaluated and compared three different approaches to calculate collisional energy loss and found that the approach proposed by Braaten and Thoma ( based on the Hard Thermal Loop (HTL) framework \cite{Braaten:1991we}) aligns better with recent experimental data. Therefore, we employ the same method for calculating collisional energy loss in our simulations.
Radiative energy loss is calculated using the reaction operator formalism (DGLV) \cite{Gyulassy:2000fs, Djordjevic:2003zk, Wicks:2005gt} and the generalized dead cone approach \cite{Saraswat:2015yll, Saraswat:2017vuy}.
\\

\begin{figure}[htbp!]
	\begin{center}
		\vspace{0.4cm}
	\resizebox{0.60\textwidth}{!}{\includegraphics{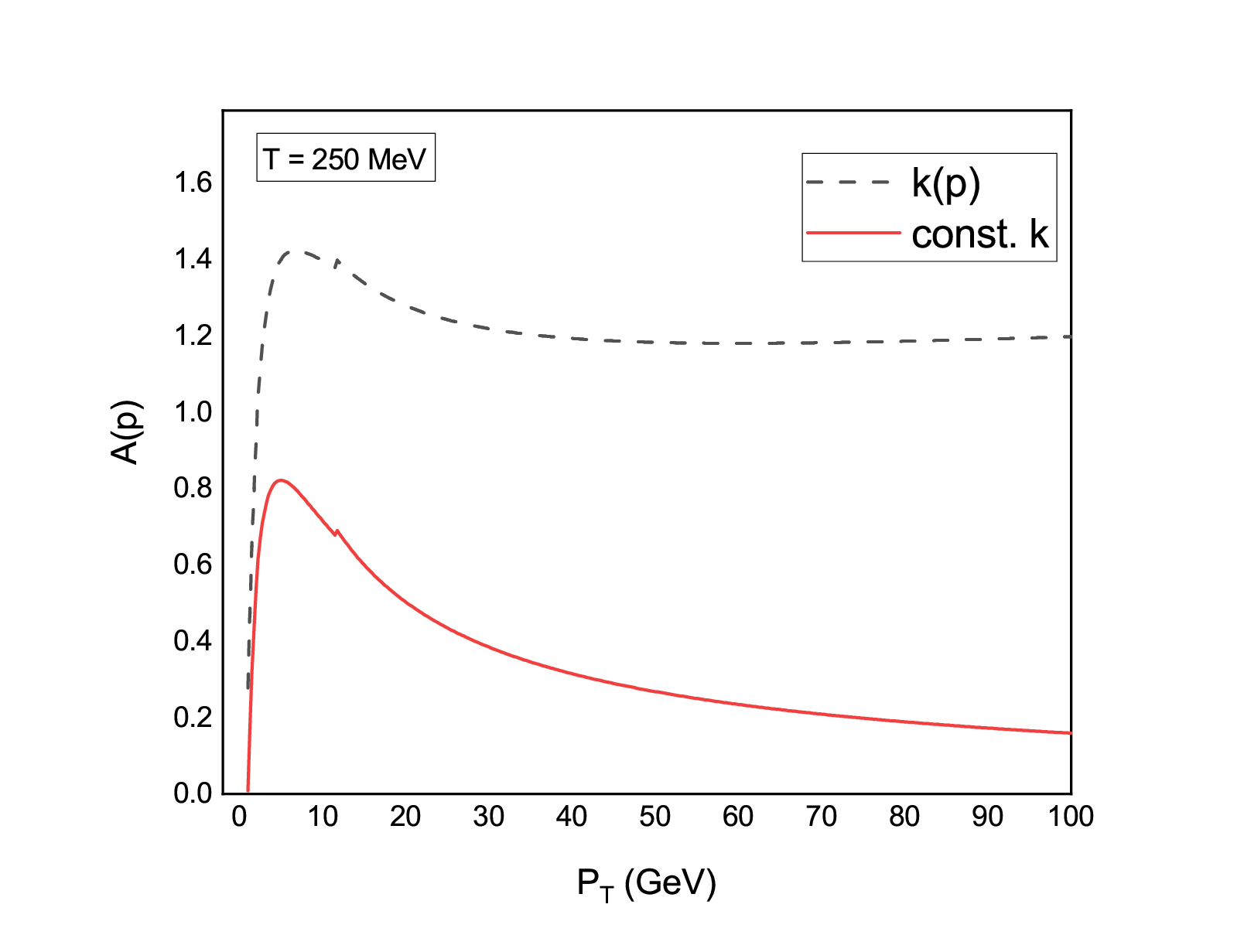}}
	\caption{\small{Drag coefficient as a function of $P_T$, calculated by constant rate (solid line) and momentum-dependent rate (dashed line) at T=250 MeV.}}\label{fig: A(p)}
\end{center}
\end{figure}

Figure (\ref{fig: A(p)}) presents the drag coefficient of charm quark for both cases of momentum-dependent coefficients and constant coefficients, at $T=250 \; MeV$. 
As shown in the figure, incorporating the linear momentum-dependent term in the drag coefficient calculation leads to a significant increase in its value at higher momenta. Consequently, greater energy suppression is expected in these regions.
By comparing the $R_{AA}$ results obtained from the above equation with those derived using constant energy loss coefficients (i.e., only the first term of the expansion), the degree of variation or improvement in the results can be quantified.
\\

\FloatBarrier
\section{Results}\label{sec:results}

The nuclear suppression factor, $R_{AA}$, defined as the ratio of quenched to unquenched transverse momentum ($p_T$) distributions, quantifies the medium-induced energy loss of heavy quarks. 
It is influenced by several key processes, including the initial production of heavy quarks, the subsequent evolution of the quark-gluon plasma, and interactions of heavy quarks within the medium. 
In this work, we focus on the momentum-dependent drag coefficient and its effect on heavy-quark energy loss, formulating our study entirely at the heavy-quark level to establish a baseline quantification of drag-momentum dependence. 
A direct hadronization model, whether via fragmentation or coalescence, is not explicitly included. Instead, we employ an effective normalization procedure, a standard approach in baseline Fokker–Planck transport analyses \cite{Jamil:2010te, Sheibani:2021ovo}, which allows qualitative comparison with D-meson $R_{AA}$ while maintaining the focus on isolating the effect of the momentum-dependent drag.
%
\\
The nuclear suppression factor can then be expressed as the ratio of the final heavy-quark distribution in heavy-ion collisions to the initial heavy-quark distribution in proton-proton collisions:
\begin{equation}
	R_{AA}(p_T) \approx \frac{f_f(P_T)^{A-A}}{f_i(P_T)^{P-P}}
\end{equation}
The proportionality can be replaced with an equality by accounting for the number of binary collisions and the relevant degrees of degeneracy. This correction is implemented through an optimization procedure in our analysis.
\\
In this section, we calculate the nuclear suppression factor of charm and bottom quarks in a Pb-Pb collision at a center-of-mass energy of 5.02 TeV and compare our findings with recent LHC experimental data collected in 2021 and 2022 \cite{ALICE:2020sjb, ALICE:2021rxa, ATLAS:2021xtw}.
The parton distribution functions are evolved using the Fokker–Planck equation, which we solve numerically on a Bjorken expanding medium, until the fireball reaches its freeze-out temperature of $T_c = 155$ MeV.
We calculate the drag coefficient using both cases of the first-order momentum expansion (Eq. \ref{bigdrag}) and the constant‑coefficient approximation (Eq. \ref{1st Drag}). This procedure allows us to compare the results with experimental data and investigate the impact of momentum dependence of energy loss coefficients on results.
\\
The final outcomes are optimized to align with experimental data by adjusting free parameters such as $\alpha$, $\beta$, $k_1$ and $k_2$, by minimizing the Chi-squared value:
\begin{equation}
	\chi^2=\sum_i \frac{ \left[ R_{AA}^{exp}  (p_T(i)) - R_{AA}^{th}  (p_T(i)) \right] ^2}{\sigma_i^2}
\end{equation}
$R_{AA}^{exp}$ and $R_{AA}^{th}$ represent experimental data and our theoretically calculated values for the suppression factor, respectively, while $\sigma$ indicates experimental error.
\\
In our parameter optimization process, we use the Minuit package \cite{Minuit.James:1975dr}, which is an effective and powerful tool that enables us to fit multiple free parameters and achieve high accuracy in minimizing the chi-squared value. This tool is essential for ensuring that the optimization is both precise and reliable, thereby enhancing the overall quality of the results.
\\
\begin{figure}[htb]
	\begin{center}
		\vspace{0.4cm}
	\resizebox{0.60\textwidth}{!}{\includegraphics{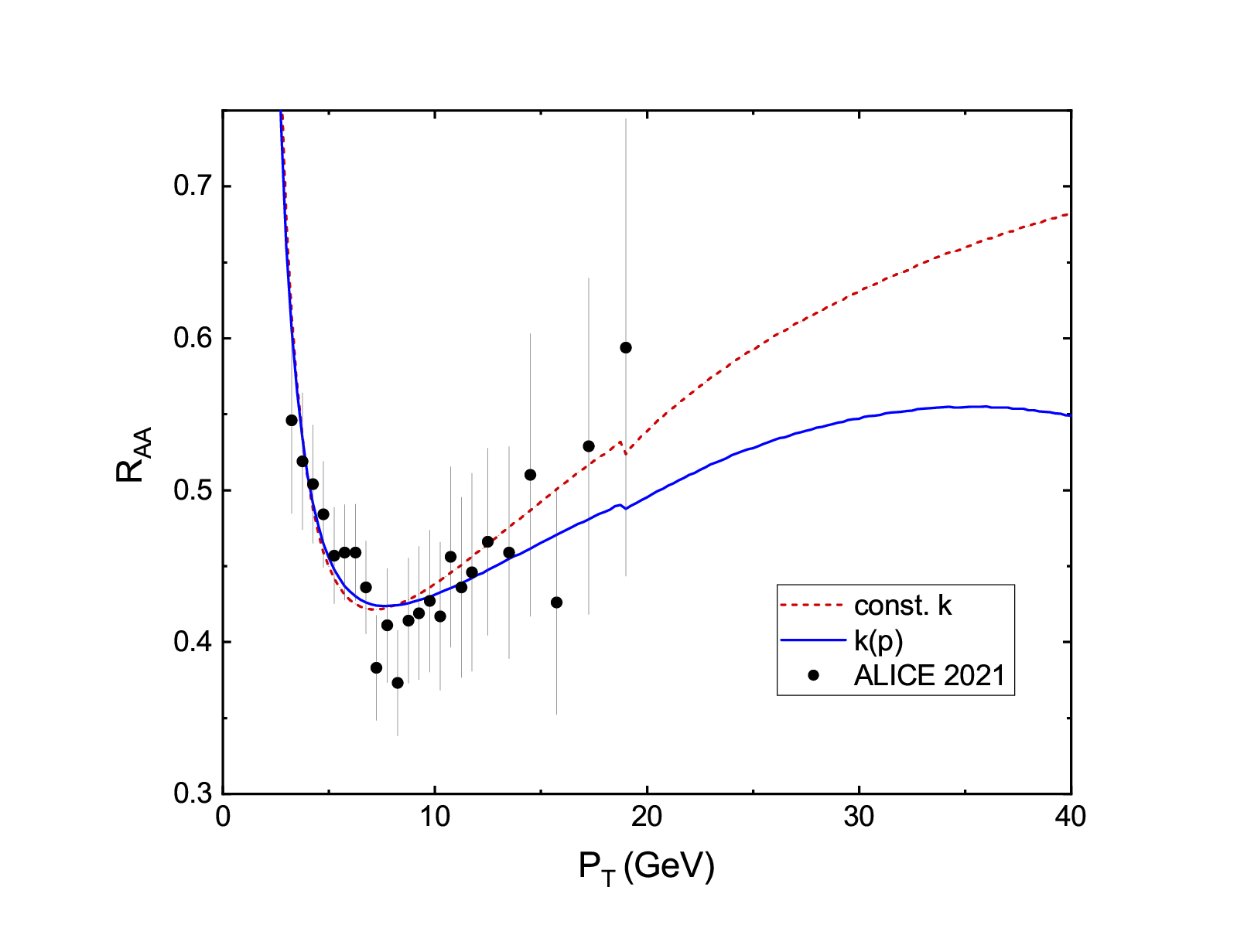}}
	\caption{\small{Nuclear modification factor of charm quark in the (20–40\%) centrality Pb-Pb collisions at $\sqrt{S_{NN}}=5.02$ TeV compared with ALICE 2021 data at forward rapidity ($2.5<y< 4$) \cite{ALICE:2020sjb}. The solid line represents the results obtained from the momentum-dependent coefficients, and dashed line corresponds to the drag coefficient with constant coefficients. }}\label{fig: c2021}
\end{center}
\end{figure}
\begin{figure}[htb]
	\begin{center}
		\vspace{0.4cm}
	\resizebox{0.60\textwidth}{!}{\includegraphics{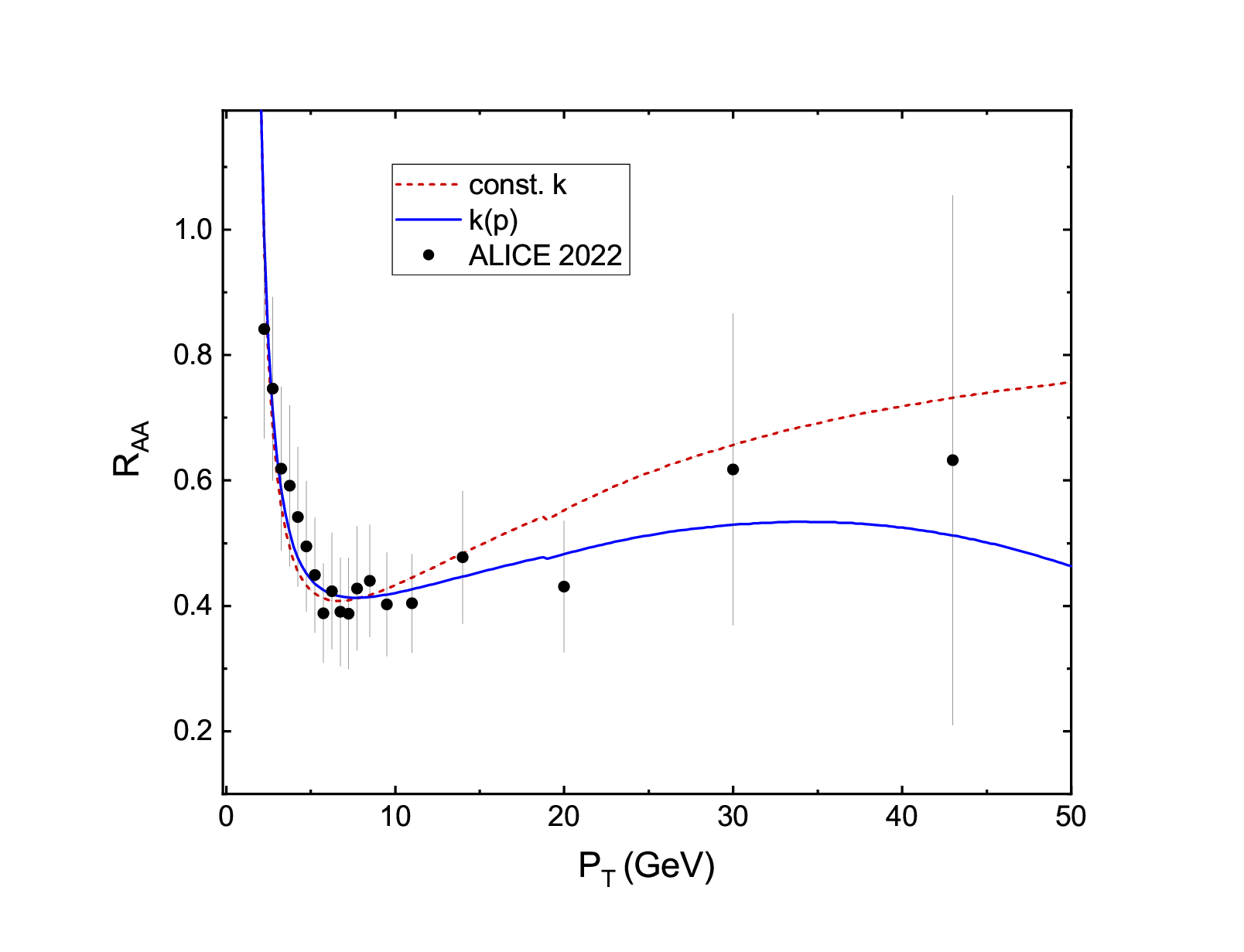}}
	\caption{\small{Nuclear modification factor of charm quark in the (30–50\%) centrality Pb-Pb collisions at $\sqrt{S_{NN}}=5.02$ TeV compared with ALICE 2022 data at mid-rapidity ($|y|<0.5$) \cite{ALICE:2021rxa}. The solid line represents the results obtained from the momentum-dependent coefficients, while the dashed line corresponds to the drag coefficient with constant coefficients. }}\label{fig: c2022}
\end{center}
\end{figure}
\\
Figures (\ref{fig: c2021}) and (\ref{fig: c2022}) demonstrate $R_{AA}$ results for charm quark, obtained by fitting theoretical predictions to the ALICE 2021 \cite{ALICE:2020sjb} and ALICE 2022 \cite{ALICE:2021rxa} datasets, respectively. 
A comparison of the two figures indicates that incorporating a momentum-dependent drag coefficient leads to better agreement with the experimental data, \textcolor{color}{particularly in the 2--15 GeV momentum region where most experimental data are located. At higher momentum, both models are statistically indistinguishable within uncertainties.}
This improvement is also supported by the $\chi^2 / dof$ value which decreases from 0.3 to 0.2 for ALICE 2022 dataset and from 0.47 to 0.44 for ALICE 2021 dataset.
\\
The coefficients $\alpha$ and $\beta$ in Equation (\ref{bigdrag}) for the charm quark are found to be $\alpha= 0.02$ and $\beta= 0.05$. 
These values indicate that both collisional and radiative energy losses of particles increase effectively at higher momentum. 
Specifically, the impact of particle's momentum on radiative energy loss becomes more significant and must be taken into account. 
We also obtain \(k_2 \approx 1.4\,k_1\) through the optimization procedure \textcolor{color}{performed over the full transverse momentum range covered by the experimental data}, indicating that the radiative energy loss of charm quarks is consistently larger than the collisional one.
\\
\begin{figure}[htb]
	\begin{center}
		\vspace{0.4cm}
	\resizebox{0.65\textwidth}{!}{\includegraphics{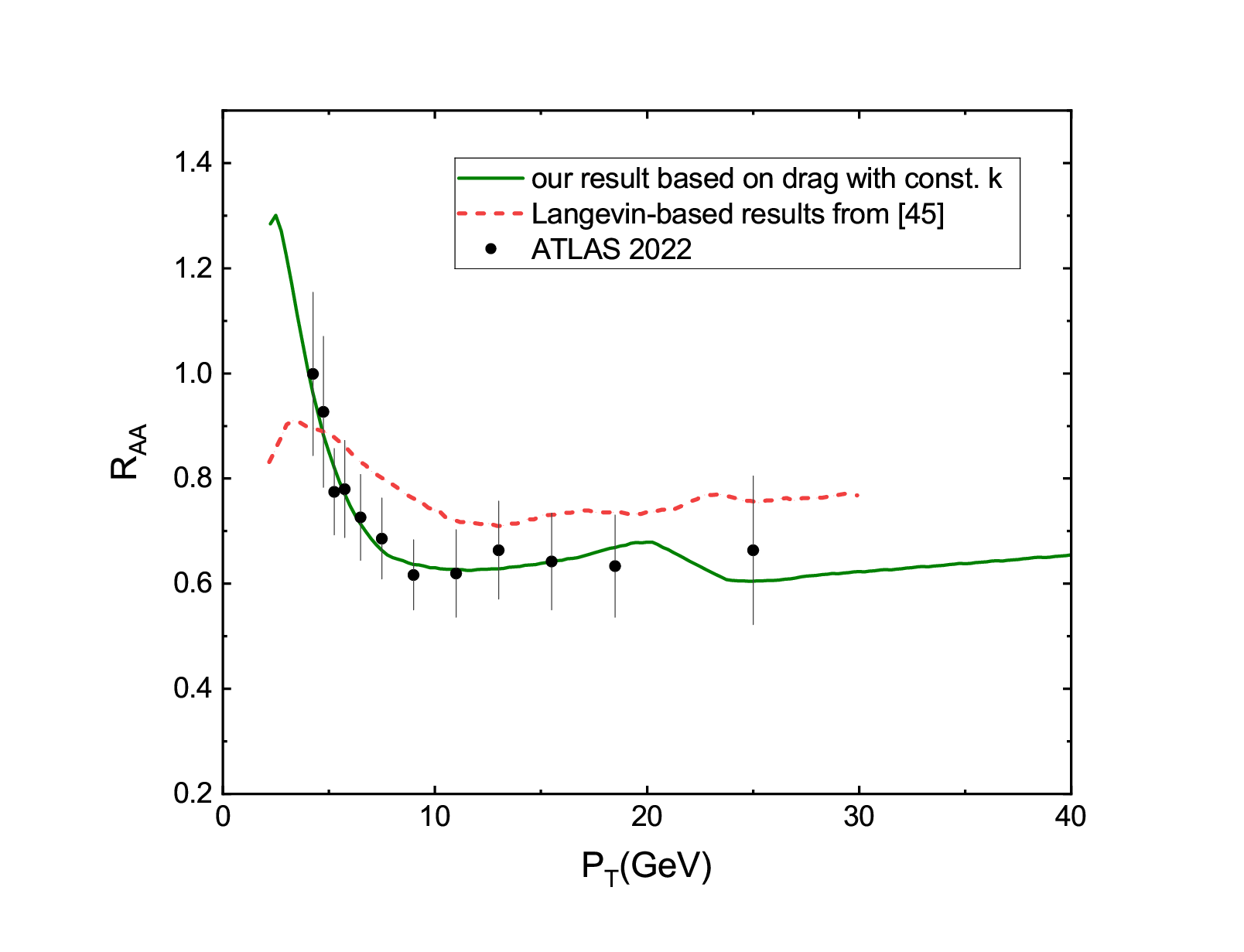}}
	\caption{\small{Nuclear modification factor of bottom quark in the (40–60\%) centrality Pb-Pb collisions at $\sqrt{S_{NN}}=5.02$ TeV (solid line) is compared with ATLAS collaboration data \cite{ATLAS:2021xtw} and theoretical predictions from Shu-Qing Li et al. \cite{Li:2024wqq} (dashed line).}}\label{fig: other model}
\end{center}
\end{figure}
Figure (\ref{fig: other model}) shows our calculated $R_{AA}$ result for bottom quark obtained by fitting to the ATLAS 2022 data \cite{ATLAS:2021xtw}.
In this analysis, we have considered only the zeroth-order terms (i.e., the constant coefficients $k_1$ and $k_2$) in the drag coefficient.
The fitted coefficients in Eq. (\ref{1st Drag}) are found to satisfy approximately $k_1 \approx 3 \; k_2$, \textcolor{color}{primarily within the momentum range covered by the available experimental data.}  

%
This hierarchy indicates a clear dominance of collisional over radiative energy loss for bottom quarks.
Also, excellent agreement between theoretical predictions and experimental data is confirmed by our $\chi^2/ dof$ value of 0.124, as illustrated in the figure (\ref{fig: other model}).
For further validation, we compare our calculated $R_{AA}$ values for bottom quark with Langevin-equation-based results reported in Ref. \cite{Li:2024wqq}, also shown in Figure (\ref{fig: other model}).
\\
Due to the limited experimental data for the bottom quark, it is not feasible to accurately fit a model with too many free parameters. 
Therefore, investigating higher-order contributions to the drag coefficient is more challenging for bottom.
Nevertheless, by repeatedly performing the fitting process,
we can constrain the approximate bounds of the free parameters.
By considering higher-order terms of drag coefficient, the parameters $\alpha$ and $\beta$ for this quark are found to be similar to those of charm quark. 
The key difference is that for bottom quarks, even by considering higher-order coefficients, we find \(k_1 > k_2\), which indicates that collisional energy loss is the dominant mechanism even when higher-order terms are included. This behavior is mainly due to the large mass of the bottom quark, which strongly suppresses gluon radiation \textcolor{color}{in the kinematic regime probed by the present data} \cite{Dokshitzer:2001zm}.\\
\begin{figure}[H]
	\begin{center}
		\vspace{0.4cm}
	\resizebox{0.65\textwidth}{!}{\includegraphics{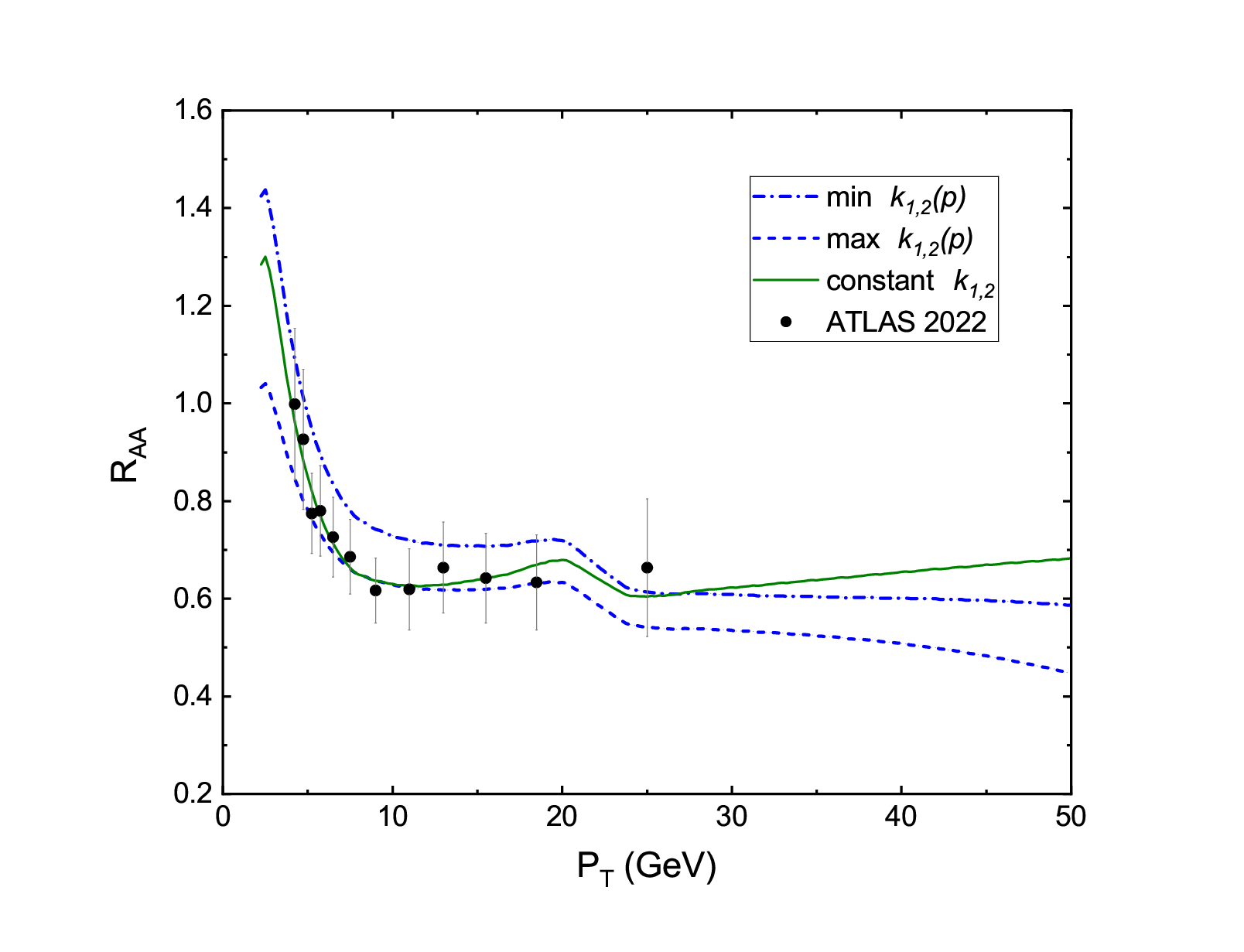}}
	\caption{\small{Nuclear modification factor of bottom quark in the (40–60\%) centrality Pb-Pb collisions at $\sqrt{S_{NN}}=5.02$ TeV compared with ATLAS 2022 data \cite{ATLAS:2021xtw}.}}\label{fig: bb}
\end{center}
\end{figure}

Figure (\ref{fig: bb}) presents the results derived from the momentum-dependent drag coefficients, for both upper and lower bounds of these coefficients, and compares them with the results based on constant coefficients. 
As shown in this figure, within the intermediate momentum range where experimental data are available, the drag coefficient with constant parameters provides satisfactory agreement with the experimental measurements.
However, at higher momenta, \textcolor{color}{where experimental data are limited and the model is less constrained,} the results obtained using momentum-dependent coefficients suggest different behaviors, indicating a stronger suppression pattern.
This observed discrepancy highlights the importance of future experimental data in the high-momentum regime to assess whether incorporating momentum dependence terms in the drag coefficient leads to more accurate theoretical predictions for bottom quarks.

\section{Summary and discussions}\label{sec: summary}

We have presented a comprehensive analysis of the drag coefficient within the framework of the Fokker–Planck equation.
The momentum dependence of the heavy-quark scattering rate has been incorporated by modeling the energy loss parameters as functions of momentum. 
The resulting drag coefficient has been compared with that obtained using a simplified model based on constant coefficients.
Our findings have indicated that considering linear momentum-dependent terms into the drag calculation has allowed for a more accurate identification of both radiative and collisional energy losses for particle distribution in momentum space. The calculated \( R_{AA} \) values have shown improved agreement with experimental data. Furthermore, the results have demonstrated that energy loss increases with particle momentum, with the enhancement being more pronounced in radiative energy loss. 
An analysis of bottom and charm quarks has revealed that radiative energy loss dominates for c quarks, whereas collisional energy loss is more significant for b quarks due to their higher mass. However, accurately determining the free parameters in the evolution of b quarks has remained challenging because of limited experimental data.
With more experiments in the future and an expanded dataset for b quark, we will be able to accurately determine the momentum-dependent coefficients. 
This will allow for a more precise quantification of the influence of heavy particle momentum on their energy loss.\\
\textcolor{color}{The present framework should be regarded as a baseline phenomenological study. The introduced momentum dependence does not aim to replace microscopic calculations, but to quantify the sensitivity of observables to effective modifications of transport coefficients. Future work including full hydrodynamic evolution and hadronization effects will further constrain these parameters.}
%


\end{document}